\documentclass[aps,twocolumn,bibnotes,preprintnumbers,longbibliography,groupedaddress]{revtex4-2} 

\usepackage[utf8]{inputenc}
\usepackage{amsmath}
\usepackage{amssymb}
\usepackage{color}
\usepackage{braket}
\usepackage{slashed}
\usepackage{xcolor}
\usepackage{ulem}
\usepackage{graphicx}
\usepackage{mathrsfs}
\usepackage{tikz}
\usepackage{subfig}
\usepackage[pdfpagelabels=false,colorlinks=true,urlcolor=teal,citecolor=teal]{hyperref}
\usepackage{mathrsfs}
\newcommand{\rd}{\mathrm{d}}

\newcommand{\nn}{\nonumber}

\newcommand{\intlim}[3]{\int_{#1}^{#2}\! \rd #3 \,}

\begin{document}
\title{Transverse-Momentum Subtraction for Semi-Inclusive Deep-Inelastic Scattering}
\author{Jun~Gao$^{1*}$,~Hai~Tao~Li$^{2*}$,~Hua~Xing~Zhu$^{3,4*}$,~Yu~Jiao~Zhu$^{5*}$}
\affiliation{
	$^1$State Key Laboratory of Dark Matter Physics, Shanghai Key Laboratory for Particle Physics and Cosmology, Key Laboratory for Particle Astrophysics and Cosmology (MOE),
School of Physics and Astronomy, Shanghai Jiao Tong University, Shanghai 200240, China\\
$^2$School of Physics, Shandong University, Jinan, Shandong 250100, China\\
$^3$School of Physics, Peking University, Beijing 100871, China\\
$^4$Center for High Energy Physics, Peking University, Beijing 100871, China\\
$^5$Max-Planck-Institut f\"{u}r Physik, Werner-Heisenberg-Institut, Boltzmannstr. 8, 85748 Garching, Germany}

\begin{abstract}
Semi-Inclusive Deep-Inelastic Scattering provides unique access to the three-dimensional momentum and spin structure of the proton, enabling precise studies of parton dynamics and hadronization in QCD. 
We present a transverse-momentum subtraction approach applied to the detected hadron that enables efficient and precise calculation of higher-order QCD corrections to identified hadron production in Semi-Inclusive Deep-Inelastic Scattering.  
We demonstrate the success of the method through a next-to-next-to-leading order QCD calculation and provide fully differential phenomenological applications, which provide important ingredients for global analyses of fragmentation functions.
Our method is applicable to next-to-next-to-next-to-leading order QCD corrections for both unpolarized and polarized semi-inclusive deep-inelastic scattering. 
\\


\end{abstract}

\maketitle

\section{\textbf{ Introduction} \label{sec:introduction}}

Quantum Chromodynamics~(QCD), after five decades of intense theoretical and experimental efforts, remains the least understood fundamental force of nature, in particular in the regime where the strong interaction becomes nonperturbative~\cite{Wilson:1974sk,tHooft:1977nqb,Alkofer:2000wg,Greensite:2011zz,Gross:2022hyw}.  The forthcoming Electron-Ion Collider (EIC) will address this frontier, serving as a precision facility to image the multidimensional structure of nucleons and nuclei~\cite{Accardi:2012qut,AbdulKhalek:2021gbh,AbdulKhalek:2022hcn,Boussarie:2023izj,Wu:2025dxg,Wen:2024cfu}. Central to this program is Semi-Inclusive Deep-Inelastic Scattering (SIDIS), $\ell+N\to \ell'+h+X$. By precision measurement for the kinematical  distributions  of the identified final-state hadron, SIDIS provides unprecedented access to the three-dimensional tomographic structure of the nucleon~\cite{Sun:2014dqm,Bacchetta:2017gcc,Scimemi:2019cmh,Bacchetta:2022awv,Bacchetta:2024qre,Moos:2025sal,Yang:2024bfz,Cammarota:2020qcw,Bacchetta:2020gko,Echevarria:2020hpy,Bury:2021sue,Boglione:2021aha,Gamberg:2022kdb,Kang:2015msa,Bhattacharya:2021twu,Horstmann:2022xkk,xue:2021svd,Barone:2009hw}, the dynamics of how quark and gluon interact with the quantum vacuum through hadronization~\cite{Gao:2025hlm,Gao:2025bko,Gao:2024dbv,AbdulKhalek:2022laj,Borsa:2022vvp,Moffat:2021dji}, and the missing contributions to the proton spin puzzle~\cite{HERMES:2004zsh,Kuhn:2008sy,Aidala:2012mv,deFlorian:2014yva,Ji:2020ena,Bacchetta:2024yzl,Yang:2024drd}. 

The unprecedented precision of the EIC calls for equally precise theoretical predictions. Significant progress has been made in recent years for fixed-order calculations for SIDIS, where Next-to-Next-to-Leading Order (NNLO) results for triple differential distribution in the Bjorken scaling variable $x$, the inelasticity $y$, and the hadron momentum fraction $z$ are now available in analytic form~\cite{Abele:2021nyo,Goyal:2023zdi,Bonino:2024qbh,Bonino:2024wgg,Goyal:2024tmo,Goyal:2024emo,Bonino:2025tnf,Haug:2025ava,Bonino:2025qta,Zhou:2025lqv,Bonino:2025bqa,Goyal:2025qyu}. Equally impressive are the developments in the precision of transverse-momentum-dependent (TMD) factorization and resummation~\cite{Caucal:2024vbv,Rodini:2023plb,Ebert:2021jhy,Jaarsma:2025ksf,Scimemi:2019gge,Rein:2022odl,Luo:2019szz,Luo:2020epw,Ebert:2020qef,Ebert:2020yqt,Zhu:2025brn,Zhu:2025ixc,Zhu:2025gts,Zhang:2024mql}, 
which are essential for describing the low transverse momentum region of SIDIS. 

In this work, we present a new method for calculating fully differential SIDIS cross sections at higher orders in QCD, based on the formalism of TMD factorization and resummation.
The result is a generalization of the $q_T$-subtraction method~\cite{Catani:2007vq,Catani:2009sm,Catani:2010en,Catani:2011qz} to SIDIS, which in principle allows the perturbative calculation to be extended to any order in QCD. 
We note a similar approach is proposed for calculations of hadron production in association with jets~\cite{Fu:2024fgj}.
As a first application, we present the NNLO calculation in a fully differential manner, using the hadron energy fraction as a concrete example. 
We demonstrate that our calculation can reproduce existing analytic results, and present new results for hadron energy fraction distributions subjecting to specific kinematic cuts. We also show that our results can be easily combined with state-of-the-art TMD resummation at Next-to-Next-to-Next-to-Next-to-Leading logarithmic (N${}^4$LL) accuracy, providing precise predictions for the transverse momentum spectrum in the full phase space. Our results are realized in a parton-level Monte Carlo program, suitable for phenomenological studies and global analyses of parton distribution functions (PDFs) and fragmentation functions (FFs) at NNLO. This formalism is readily generalizable to polarized SIDIS, as well as to N$^3$LO calculations, contingent upon the availability of the resolved NNLO corrections defined below.
 
\section{\textbf{ Methodology} \label{sec:method}}
We consider semi-inclusive DIS with  high-energy lepton beams scattering off  an unpolarized nucleon target
\begin{align} \label{eq:SIDIS}
 \ell(p_\ell) + N(P_N) \to \ell'(p_{\ell'}) + h(P_h) + X(P_X)
\,,\end{align}
where the leptonic momentum transfer is defined as  $q=p_l-p_{\ell'}$, with virtuality $Q^2=-q^2$.
The standard kinematic invariants are
\begin{align}
x=\frac{Q^2}{2 P_N\cdot q}\,\quad y=\frac{P_N\cdot q}{P_N\cdot p_\ell}\,,\quad z=\frac{P_N\cdot P_h}{P_N\cdot q}\,.
\end{align}
The transverse momentum $\vec P_h^\perp$ of the detected hadron is defined following the Trento conventions~\cite{Bacchetta:2004jz}, where the transverse plane is orthogonal to the directions spanned by the virtual photon and the incoming nucleon. The corresponding transverse metric is
\begin{align}
[g_{\perp}^{\mu \nu}]_\text{Trento}=g^{\mu\nu}-\frac{q^\mu P_N^\nu+P_N^\mu q^\nu}{P_N\cdot q}-\frac{Q^2P_N^\mu P_N^\nu}{(P_N\cdot q)^2}\,.
\end{align}
In this convention, at leading order (LO) the transverse momentum of the virtual photon in the hadron frame is related to the hadron transverse momentum by
$\vec q_\perp=-\vec P^\perp_h/z $.
The differential SIDIS cross section can be decomposed into two complementary regions
\begin{equation}
\label{eq:slicing}
 \frac{d\sigma}{d\mathcal{O}}  = \int_{q_{\rm T, cut}}^{q_{\rm T, max}} d q_T \frac{d\sigma}{d q_T {d\mathcal{O}} } +\int_{0}^{q_{\rm T, cut}} d q_T \frac{d\sigma}{d q_T {d\mathcal{O}} } 
 \,,
\end{equation}
where the first term corresponds to the large-$q_T$
 (resolved) region, which is described by collinear factorization and fixed-order perturbation theory. 
 It can be calculated using local subtraction methods developed for identified hadron production at NLO~\cite{Daleo:2004pn,Kniehl:2004hf,Wang:2019bvb,Liu:2023fsq,Zidi:2024lid,Caletti:2024xaw,Bonino:2024adk} and at NNLO~\cite{Czakon:2025yti,Generet:2025bqx}.
 Here ${q_{\rm T, cut}}$ is a resolution/slicing parameter introduced in the $q_T$-subtraction method~\cite{Catani:2007vq}.
The low-transverse-momentum (unresolved) part is governed by the leading-power  TMD factorization formula~\cite{Meng:1995yn,Nadolsky:1999kb,Ji:2004wu,Ji:2004xq,Koike:2006fn,Collins:2011zzd}, which resums logarithmically enhanced singular contributions at low transverse momentum and expresses the cross section in terms of universal TMD parton distribution  and fragmentation functions 
\begin{widetext}
\begin{align}
\label{eq:SIDIS-fac-semi-CS}
\frac{1}{\sigma_0}\; \frac{\rd \sigma_{\ell+N\to \ell'+h+X}}{\rd^{2} \vec q_\perp \rd x\rd y\rd z}
\simeq&\,
\sum_q   H_q(Q^2,\mu) \times
\int \frac{\rd^2 \vec b_\perp}{(2\pi)^2} e^{i \vec b_\perp \cdot\vec q_\perp}\, 
 f_1^q(x,b_\perp,\xi^{n}_0,\mu^n_0) \times 
 D_1^{q}\left(z, \frac{b_\perp}{z},\xi^{\bar n}_0,\mu^{\bar n}_0\right)
\nn\\
\times&
\,\prod_ie^{-2 K^i_{\rm cusp}(\mu^i_0,\mu)+A^i_H(\mu^i_0,\mu)}
\times
\left(
\frac{\xi^i}{{\mu^i_0}^2}
\right)^
{
A^i_{\rm cusp}(\mu^i_0,\mu)
}
\times
\left(\frac{\sqrt{\xi^i}}{\sqrt{\xi^i_0}}\right)
^
{
K^i(b_\perp,\mu_0^i)
}
\,,
\end{align}
\end{widetext}
where $\alpha=e^2/(4 \pi)$  is the  electromagnetic coupling  and
$H_q(Q^2,\mu)$ represents the  square of the hard matching coefficient~\cite{Becher:2006mr,Moch:2005tm,Moch:2005id,Baikov:2009bg,Gehrmann:2010ue,Gehrmann:2010tu,Lee:2022nhh}.
The born-level cross-section is 
\begin{align}
\sigma_0=2\pi \alpha^2 \frac{1+(1-y)^2}{ Q^2 y} \,.
\end{align}
The functions 
$f_1^q$
and 
$D_1^{q}$
denote the TMD parton distribution and fragmentation functions, respectively. 
In addition to the renormalization scale $\mu$, they depend on rapidity scales introduced through the Collins-Soper (CS) formalism~\cite{Collins:1981uk,Chiu:2012ir,Chiu:2011qc}. 
For SIDIS, a canonical choice of the CS scales is
\begin{align}
\label{eq:CS-scale}
\xi^n = Q^2 \frac{x}{x+y-x y}\,,\quad \xi^{\bar n} = Q^2 \frac{x+y-x y}{x}\,.
\end{align}
The rapidity evolution with respect to the CS scales is controlled by the CS kernel, 
\begin{align}
 K^i(b_\perp,\mu)=   -2A^i_{\rm cusp}(\mu,\mu_b)
+
\gamma_R^i(\mu_b)\,,
\end{align}
which depends on the impact-space parameter $b_T=|\vec b_\perp|$ through scale $\mu_b=b_0/b_T$, with $ b_0=2e^{-\gamma_E}$.
The CS kernel includes a perturbative small-$b_T$ contribution, 
its ingredient $\gamma_R$, known as the QCD rapidity anomalous dimension, has reached N${}^4$LL accuracy~\cite{Li:2016ctv,Moult:2022xzt,Duhr:2022yyp}.
Nonperturbative corrections at large-$b_T$ can be obtained from recent lattice QCD advances~\cite{LatticePartonLPC:2022eev,Shanahan:2020zxr,Shanahan:2021tst,Avkhadiev:2024mgd,Schlemmer:2021aij,Li:2021wvl,LatticeParton:2020uhz,Tan:2025ofx}.
The remaining Sudakov ingredients $A_{\gamma}$, and $K_{\rm cusp}$~\cite{Collins:1984kg,Becher:2006mr,Collins:2011zzd}
\begin{align} \label{eq:Kw_def}
&K_\Gamma (\mu_0, \mu)
=- \intlim{\alpha_s(\mu_0)}{\alpha_s(\mu)}{\alpha_s} \frac{\Gamma(\alpha_s)}{\beta(\alpha_s)}
   \intlim{\alpha_s(\mu_0)}{\alpha_s}{\alpha_s'} \frac{1}{\beta(\alpha_s')}
\,,
\nn\\
&A_\gamma(\mu_0, \mu)
= -\intlim{\alpha_s(\mu_0)}{\alpha_s(\mu)}{\alpha_s} \frac{\gamma(\alpha_s)}{\beta(\alpha_s)}
\,,\end{align}
controlled by the $\beta$-function~\cite{Herzog:2017ohr,vanRitbergen:1997va,Czakon:2004bu,Baikov:2016tgj,Luthe:2017ttg,Chetyrkin:2017bjc}, the cusp and virtual anomalous dimensions, are also known to high perturbative accuracy~\cite{Grozin:2015kna,Henn:2016men,vonManteuffel:2016xki,Henn:2016wlm,Lee:2017mip,Herzog:2018kwj,Lee:2019zop,vonManteuffel:2019wbj,Bruser:2019auj,Das:2019btv,vonManteuffel:2020vjv,Agarwal:2021zft,Lee:2021uqq,Lee:2022nhh,Chakraborty:2022yan}.

When $\Lambda_{\rm{QCD}}\ll q_T \ll Q$, the singular terms in  Eq.~(\ref{eq:slicing}) are reproduced by TMD factorization formula Eq.~(\ref{eq:SIDIS-fac-semi-CS}) through twist-2 matchings of the TMDs~\cite{Luo:2019szz,Luo:2020epw,Ebert:2020qef,Ebert:2020yqt}, up to nonsingular power corrections.
Importantly, all ingredients needed for a calculation at ${\rm N^3LO}$ for SIDIS are available by building from the results above. 
\section{\textbf{ Applications} \label{sec:app}}
\begin{figure}[t]
    \includegraphics[width=1.0\linewidth]{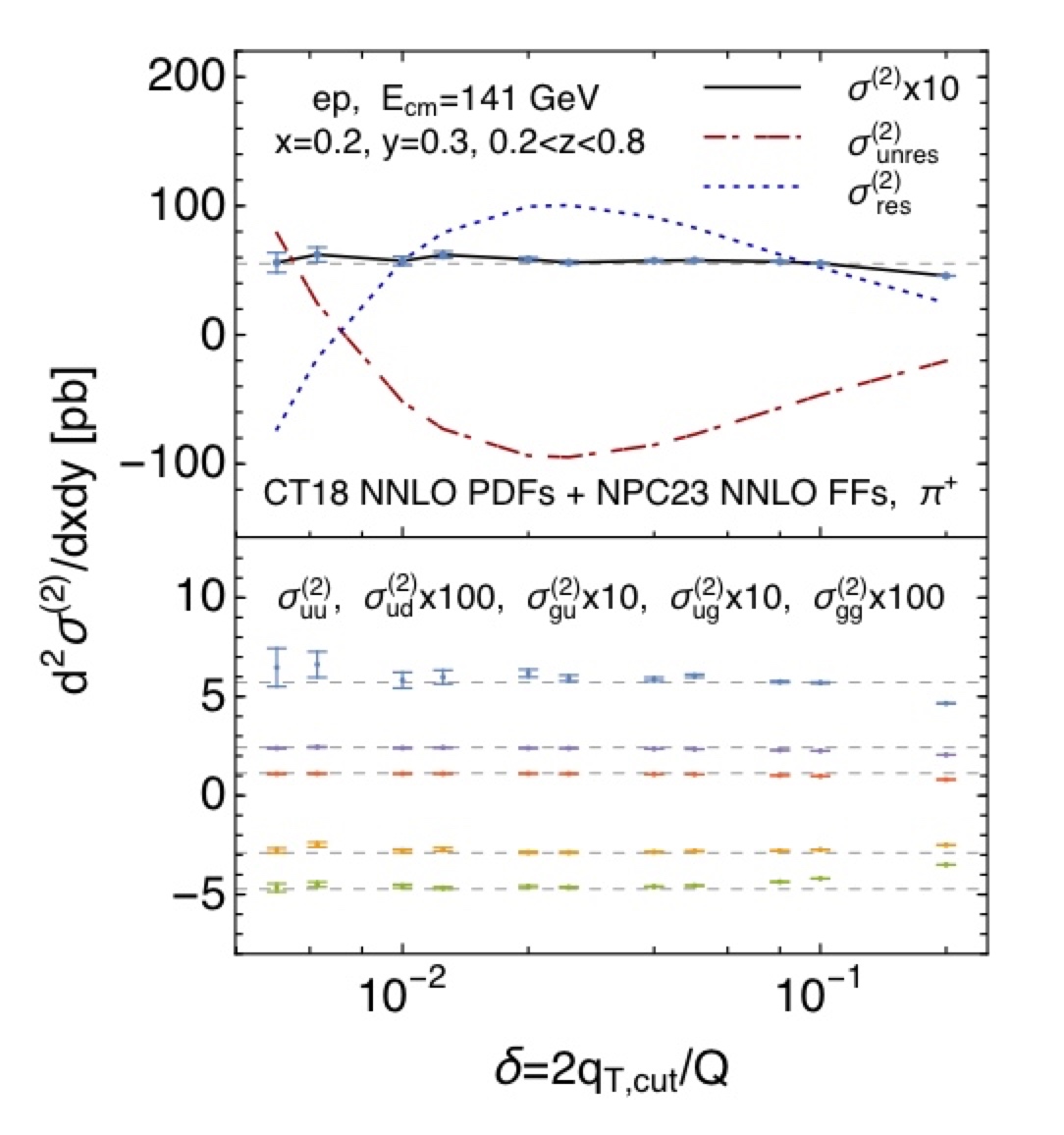}
    \caption{%
    NNLO QCD corrections to the differential cross section at the EIC for charged pion production, with $x=0.2$, $y=0.3$, and $0.2<z<0.8$,  as a function of the slicing parameter. The resolved, unresolved, and full contributions are represented by the dotted, dash-dotted, and solid lines, respectively.
    The dashed line is analytical calculations for different partonic channels~\cite{Bonino:2024qbh,Goyal:2024tmo}.
    The lower panel shows the cross sections for different partonic channels.
    }
    \label{fig:fig1}
\end{figure}

\begin{figure}[t]
    \includegraphics[width=1.0\linewidth]{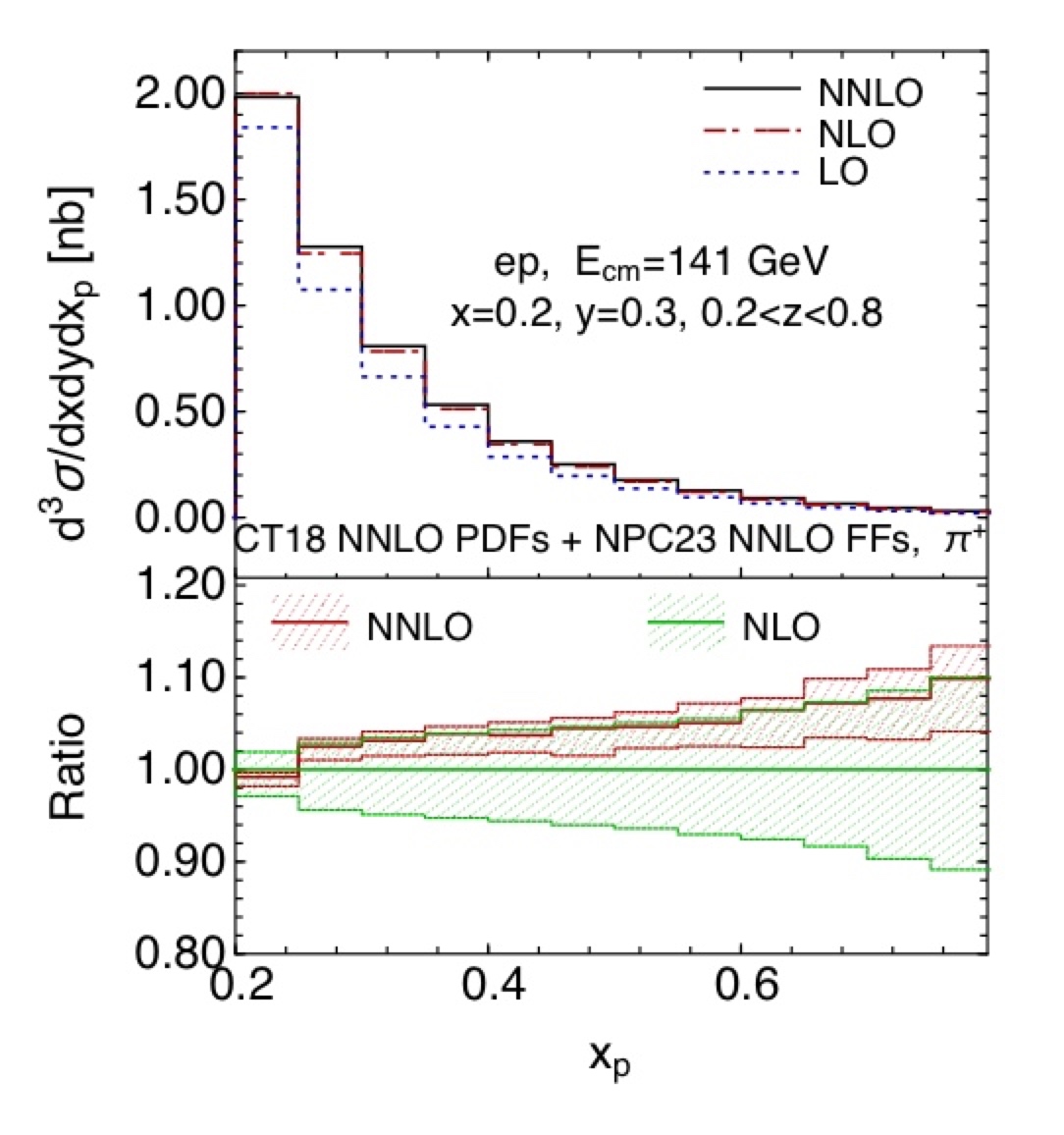}
    \caption{%
    Differential cross section as functions of the hadron energy fraction $x_p$ for charged pion production at the EIC at various orders in QCD together with scale variations.
    }
    \label{fig:fig2}
\end{figure}

\begin{figure}[t]
    \includegraphics[width=1.0\linewidth]{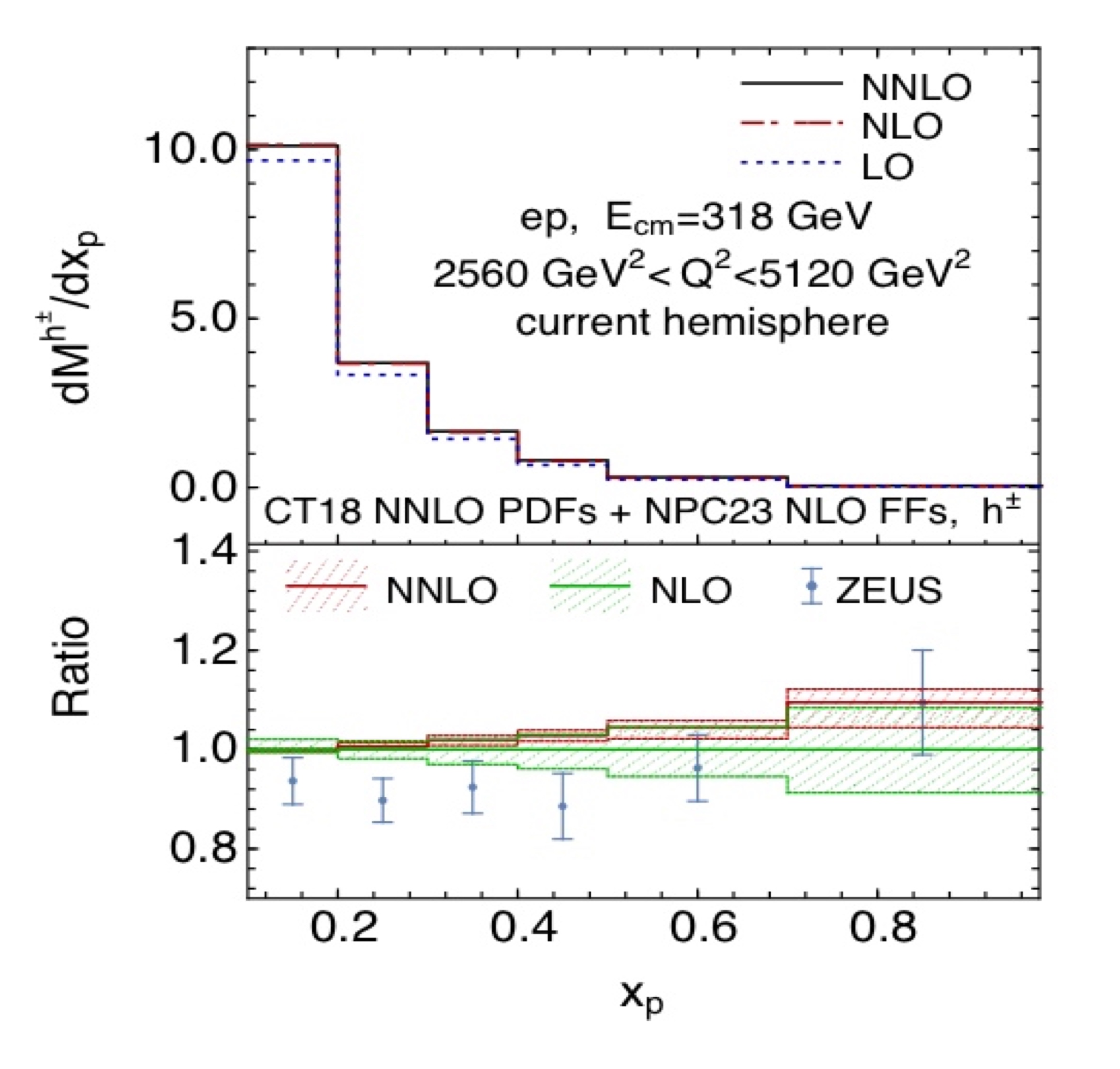}
    \caption{%
    Multiplicity distribution as functions of the hadron energy fraction $x_p$ for unidentified charged hadron production at the HERA at various orders in QCD together with scale variations, compared to the ZEUS data~\cite{ZEUS:2010mrq}.
    }
    \label{fig:fig3}
\end{figure}

In this section we present phenomenological studies of SIDIS at NNLO in QCD.
We consider electron-proton collisions at center of mass energies of the EIC ($\sqrt s=141$ GeV) or HERA ($\sqrt s=318$ GeV), and use the CT18 NNLO  PDFs~\cite{Hou:2019efy} of 5 active quark flavors with the strong coupling constant $\alpha_S(M_Z)=0.118$.  
We calculate differential cross sections for charged pion ($\pi^+$) and unidentified charged hadrons ($h^{\pm}$) production.
In the case of charged pion we use the NPC23 NNLO FFs~\cite{Gao:2025hlm} through all orders, while only NLO FFs are available for the unidentified charged hadrons~\cite{Gao:2024nkz,Gao:2024dbv}.  
We have neglected contributions from the $Z$ boson exchange for simplicity, which are small for the energies considered.
The contribution from the resolved part is calculated with the {\tt FMNLO} program based on a hybrid subtraction scheme~\cite{Liu:2023fsq}.
In Fig.~\ref{fig:fig1} we show the NNLO corrections to the differential cross section of charged pion production at the EIC. %
We fix the kinematic variables of Bjorken $x=0.2$ and inelasticity $y=0.3$ while integrating over the hadron light-cone momentum fraction $0.2<z<0.8$.
The upper panel presents the resolved (dotted line) and unresolved (dash-dotted line) results from the subtraction method for different choices of the slicing parameter $\delta \equiv 2q_{\rm T, cut}/Q$.
The full results, which are the sum of the resolved and unresolved, together with uncertainties from MC integrations, are indicated by the error bars joined with a solid line.
We have multiplied the full results by a factor of ten for convenience, and compared them to those from an independent calculation based on the analytical results in Refs.~\cite{Bonino:2024qbh,Goyal:2024tmo}, represented by the horizontal dashed line.
Excellent agreement with the analytical results is observed, with rapid convergence for $\delta \lesssim 0.1$.
In particular, choosing the slicing parameter of $\delta \lesssim 0.05$ is sufficient to suppress power corrections to a negligible level.
In the lower panel we further investigate the NNLO corrections from different partonic channels, e.g., with the incoming parton being either a $u$ quark or gluon, and the fragmenting parton being $u$, $d$ quarks or a gluon.
The error bars from top to bottom show the contributions of the individual partonic channels, rescaled by the indicated factors for visibility,
and the corresponding dashed lines are predictions from the analytical calculations.
In all partonic channels, we find very good agreement between the subtraction method and the analytical calculations.
Our calculation is fully differential in the final-state hadron kinematics. Here, we consider the distribution of the hadron energy fraction in the Breit frame, $x_p\equiv 2E_h/Q$, called scaled momentum spectra. Importantly, beyond LO $x_p$ no longer coincides with the fragmentation variable $z$. To our knowledge, a fully differential NNLO description of $x_p$ distribution has not been available before this work.
In Fig.~\ref{fig:fig2} we compare the NNLO predictions to those at LO and NLO.
The hatched bands represent QCD scale variations calculated by varying $\mu_D$ and $\mu_R=\mu_F$ independently by a factor of two from their nominal values of $Q$~\cite{Abele:2021nyo}. 
The NNLO corrections are small, which is negligible at low $x_p$ and increases from 2\% to 10\% with increasing $x_p$.
The scale variations are suppressed by at least a factor of two across the full range of $x_p$.
Studying $x_p$ in the current hemisphere isolates the current-fragmentation region, where hadron production is dominated by the fragmentation of the struck parton and is largely insensitive to the proton remnant. This makes $x_p$ distributions a clean probe of parton-to-hadron fragmentation dynamics and a powerful testing ground for fragmentation models and perturbative QCD evolution.
For instance the ZEUS collaboration measured multiplicities of unidentified charged hadrons differential in $x_p$ in the current hemisphere~\cite{ZEUS:2010mrq}.
That selects only charged hadrons with a longitudinal momentum in the same direction as the virtual photon in the Breit frame.
The ZEUS data~\cite{ZEUS:2010mrq} and the corresponding data from H1~\cite{H1:2009lef} were included in the recent global analysis of fragmentation functions at NLO in QCD~\cite{Gao:2024dbv}, since fully differential NNLO predictions were not available at that time.
Note that the NLO FFs from~\cite{Gao:2024dbv} will be used in the calculations in Fig.~\ref{fig:fig3}, where we compare our predictions to the ZEUS data~\cite{ZEUS:2010mrq} measured in the $Q^2$ region of 2560 to 5120 ${\rm GeV}^2$ in Fig.~\ref{fig:fig3}.
The error bars denote the central values of the measurement together with total experimental uncertainties.
The NNLO corrections become more pronounced at larger values of $x_p$. However, they significantly reduce the scale uncertainties, bringing them well below the experimental uncertainties.
Therefore, our results make it possible to incorporate high-energy SIDIS data from HERA directly into global fragmentation-function analyses at NNLO in QCD.  

\begin{figure}[t]
    \includegraphics[width=1.0\linewidth]{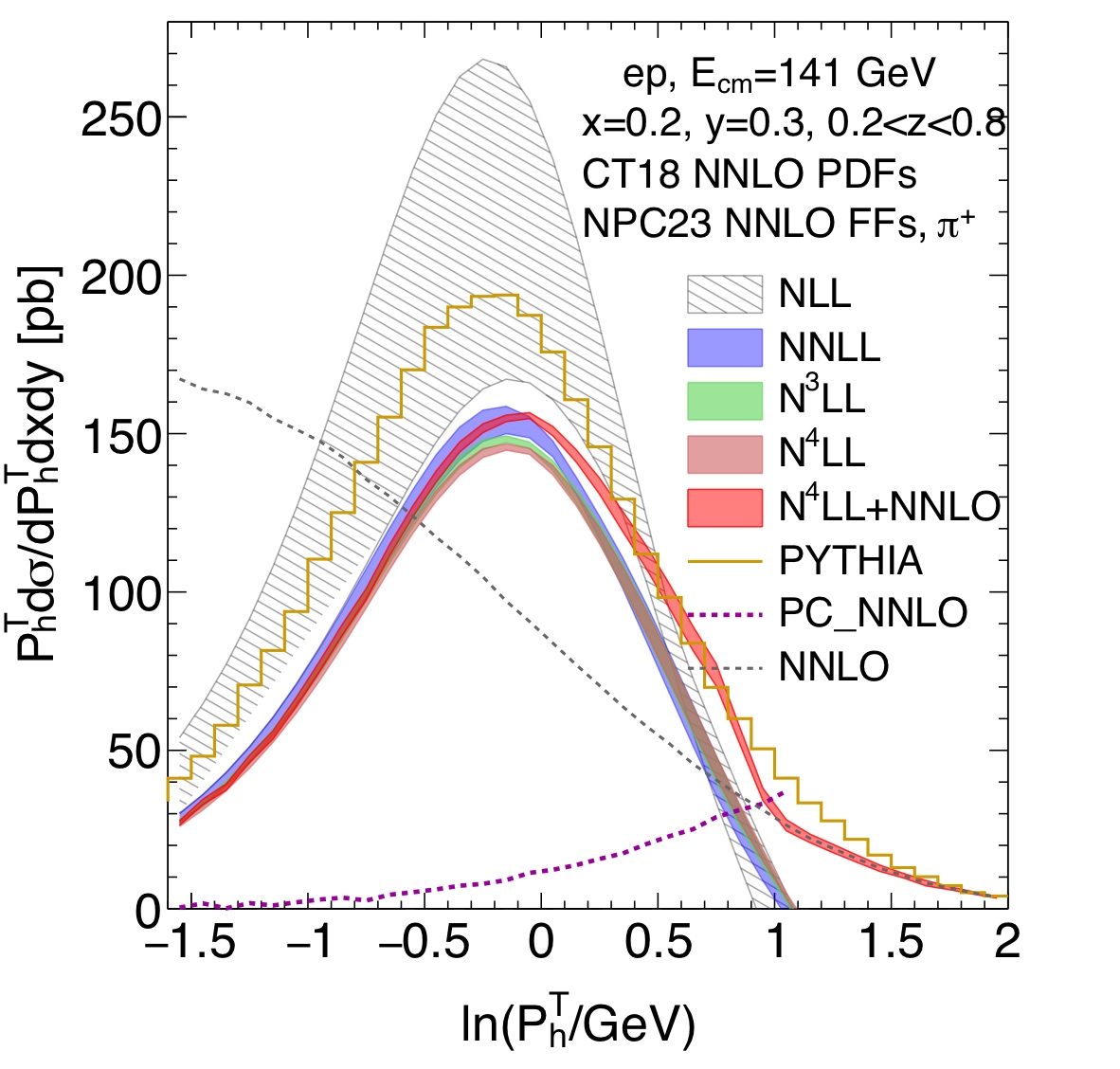}
    \caption{
    Transverse-momentum  $P_h^T$ distribution of the charged pion at the EIC with $\sqrt{s}=141$ GeV. The bands indicate the scale uncertainties.  
    The orange histogram represents the results from {\tt PYTHIA}.  
    The red band denotes the N$^4$LL-resummed result matched with NNLO fixed-order prediction. 
    The purple and gray dashed lines show the NNLO power correction and the full NNLO prediction, respectively.}
    \label{fig:ffs}
\end{figure}
The fixed-order calculation becomes unreliable when $\Lambda_{\rm{QCD}}\ll P_h^T \ll Q$, 
due to the presence of large logarithmic contributions.
In this region, the perturbative series must be resummed to all orders.
This resummation is achieved using the TMD factorization formula in Eq.~(\ref{eq:SIDIS-fac-semi-CS}), 
enabled by the twist-2 matching of TMDs~\cite{Luo:2019szz,Luo:2020epw,Ebert:2020qef,Ebert:2020yqt} 
and precise knowledge of the relevant anomalous dimensions governing the evolution factors in Eq.~(\ref{eq:Kw_def}).
In the intrinsic nonperturbative region, where $q_T\sim \Lambda_{\rm{QCD}}$, 
we adopt the BLNY18 parametrization~\cite{Sun:2014dqm}, implemented with the  $b_\ast$ prescription~\cite{Collins:2014jpa}. 
We further present in Fig.~\ref{fig:ffs} the transverse-momentum spectrum of $\pi^+$ for EIC kinematics 
at $\sqrt{s}=141$ GeV with $x=0.2$, $y=0.3$, and  hadron momentum fraction integrated over the range $0.2<z<0.8$.
The bands indicate scale uncertainties, obtained by varying the boundary scales $(\mu_0\,,\sqrt{\xi_0})$ in Eq.~(\ref{eq:SIDIS-fac-semi-CS})  
by a factor of two from their canonical values.
The Sudakov peak is at around $P_h^T=1$ GeV and is slightly shifted to the right from NLL to N$^4$LL.
The logarithmic resummation beyond NLL yields sizable corrections
and significantly reduces both the cross section and the scale uncertainties. 
The orange line represents the results from {\tt PYTHIA 8.3}~\cite{Bierlich:2022pfr}, which is consistent with the NLL result in the Sudakov region. 

The red band represents the N$^4$LL resummed result, matched to the NNLO fixed-order (up to $\mathcal O(\alpha_s^2)$) predictions.
The matched cross section is defined as
\begin{equation}
     \rd \sigma_\textrm{NNLO+$\text{N}^{4}$LL}=(1-f)\times \rd \sigma_\textrm{NNLO}+f\times \rd \sigma_\textrm{$\text{N}^{4}$LL+NS}\,,
\end{equation}
where $\rd \sigma_\textrm{$\text{N}^{4}$LL+NS}$ is given by the sum of pure N$^4$LL-resummed contribution and the nonsingular power correction,
the latter being obtained as the difference between the fixed-order prediction and the leading-power singular terms given by the TMD factorization in Eq.~(\ref{eq:SIDIS-fac-semi-CS}).
The transition function $f$ is chosen to be monotonically decreasing from $1$ to $0$ over a matching region $[a,b]$, implemented using the built-in {\tt SmoothStep} function in {\tt Mathematica}~\cite{Mathematica14.3}.
Explicitly,
\begin{equation}
f(t) =
\begin{cases}
1, & t \le a, \\[4pt]
1 - \left[ 3\left(\dfrac{t-a}{b-a}\right)^2 - 2\left(\dfrac{t-a}{b-a}\right)^3 \right], & t \in (a,b), \\[8pt]
0, & t \ge b .
\end{cases}
\end{equation}
Here $t = P_h^T/\mathrm{GeV}$, and we choose the matching region as
$2~\mathrm{GeV} \lesssim P_h^T \lesssim 3~\mathrm{GeV}$.
We find that for $P_h^T > 1$GeV, the power corrections are sizable.
For $P_h^T > 3$GeV, the resummed result turns negative and unphysical, where the resummation is no longer reliable, and the fixed-order cross section should be used instead. The difference between N${}^4$LL+NNLO and PYTHIA results at $P_h^T > 3$ GeV arises from the NNLO corrections, which are treated differently in the PYTHIA parton shower.
Our observations highlight the inclusion of higher-order radiative corrections for TMD fit at the EIC, 
both in the Sudakov region (extending from NLL to N${}^4$LL accuracy) and in the transition region where power corrections become important.

%
\section{\textbf{ Summary and Outlook} \label{sec:conclusion}}
In summary we introduce a novel framework that generalizes the $q_T$-subtraction method to the identified hadron production, enabling the calculation of QCD corrections to fully differential SIDIS cross sections. 
We validate this framework by applying it to the calculation of NNLO QCD corrections, finding excellent agreement with established analytical results. 
The fully differential results allow direct comparison with available experimental measurements, and provide new inputs for global analysis of both collinear and TMD FFs and PDFs with high precision. 
Importantly, our method can be used to compute N$^3$LO QCD corrections for both polarized and unpolarized SIDIS.
Thus it enables the development of new theoretical tools for the study of the three-dimensional internal structure of nucleons in the EIC era with unprecedented precision.
\textbf{ Acknowledgments}
The work of JG is supported by the National Natural Science Foundation of China (NSFC) under Grant No. 12275173 and the Shanghai Municipal Education Commission under Grant No. 2024AIZD007.
HTL is supported by the National Natural Science Foundation of China under grant No.  12275156, No. 12321005 and the Shandong Provincial Department of Science and Technology under Project No. TQ012025001. 
HXZ is supported by the National Natural Science Foundation of China under grant No. 12425505. H.X.Z. would also like to thank the Southern Center for Nuclear-Science Theory (SCNT), Institute
of Modern Physics, Chinese Academy of Sciences for hospitality while this work was carried out.
The computations in this paper were run on the Siyuan-1 cluster supported by the Center for High Performance Computing at Shanghai Jiao Tong University.

\textbf{ Conflict of interest} The authors declare that they have no conflict of interest.

\bibliographystyle{apsrev4-1} 
\bibliography{sidis}
\end{document}